\documentclass[aps,pre,onecolumn,amsmath,amssymb,10pt]{revtex4-1}

\usepackage{amssymb,amsfonts,amsmath,graphicx}
\usepackage{color}
\usepackage{natbib}
\usepackage{xcolor}
\usepackage{float}
\usepackage{bm}
\usepackage{hyperref}
\usepackage{comment}
\usepackage{graphicx}
\usepackage{bbold}
\usepackage{subfig}
\usepackage[font=small,labelfont=bf,justification=RaggedRight,format=plain, singlelinecheck=false]{caption}
\def\be{\begin{equation}}
\def\ee{\end{equation}}
\def\bea{\begin{eqnarray}}
\def\eea{\end{eqnarray}}
\def\n{{\bf n}}

\def\be{\begin{equation}}
\def\ee{\end{equation}}
\def\bea{\begin{eqnarray}}
\def\eea{\end{eqnarray}}

\def\pt{\partial_t}

\def\a{\alpha}
\def\b{\beta}

\def\n{{\bf n}}

\def\v{{\bf v}}

\def\bs{\boldsymbol}

\usepackage[
margin=0.7in,
]{geometry}
\begin{document}

\title{Active Segregation Dynamics in the Living Cell}

\author{Ajay Bansal}
\author{Amit Das$^{\dagger}$}
\author{Madan Rao}
\email{madan@ncbs.res.in}

\affiliation{
Simons Centre for the Study of Living Machines, National Centre for Biological Sciences (TIFR), Bellary Road, Bangalore 560065, India \\
$^\dagger~$Present address: Department of Physics, Northeastern University, MA 02115, USA
}

\begin{abstract}
In this paper, we bring together our efforts in identifying and understanding nonequilibrium phase segregation driven by active processes in the living cell, with special focus on the segregation of cell membrane components driven by active contractile stresses arising from cortical actomyosin.
 This also has implications for active segregation dynamics in membraneless regions within the cytoplasm and nucleus (3d).   We formulate an active version of the Flory-Huggins theory that incorporates a contribution from fluctuating active stresses. Apart from knitting together some of our past theoretical work in a comprehensive narrative, we
 highlight some new results, and establish a correspondence with recent studies on Active Model B/B$+$.  We point to the many unusual aspects of the dynamics of active phase segregation, such as (i) anomalous growth dynamics, 
 (ii) coarsening accompanied by propulsion and coalescence of domains that exhibit nonreciprocal effects,
 (iii) segregation into mesoscale domains, (iv)  emergence of a nonequilibrium phase segregated steady state
characterised by strong macroscopic fluctuations (fluctuation dominated phase ordering (FDPO)),  and 
(v) mesoscale segregation even above the equilibrium $T_c$. 
Apart from its implications for actively driven segregation of binary fluids, 
these ideas are at the heart of an {\it Active Emulsion} description of the lateral 
 organisation of molecules on the plasma membrane of living cells, whose full molecular elaboration appears elsewhere.
\end{abstract}

\maketitle

\section{Introduction}
\label{sect:Intro}
The cell is an organised collection of interacting biomolecules and ions in a highly viscous aqueous medium that is maintained 
out-of-equilibrium~\cite{Phillips} by active stresses and fluxes~\cite{Marchetti2013}.
The active forces that drive this organisation, maintain the cellular system in a nonequilibrium steady state.
This molecular organisation and segregation at different scales is crucial for cellular function and to all processes of life.
Here, 
we will discuss 
specific nonequilibrium mechanisms governing the dynamics of segregation of molecular composition within the cell.

\begin{figure}[h]
 \centering
  \includegraphics[width=.7\linewidth]{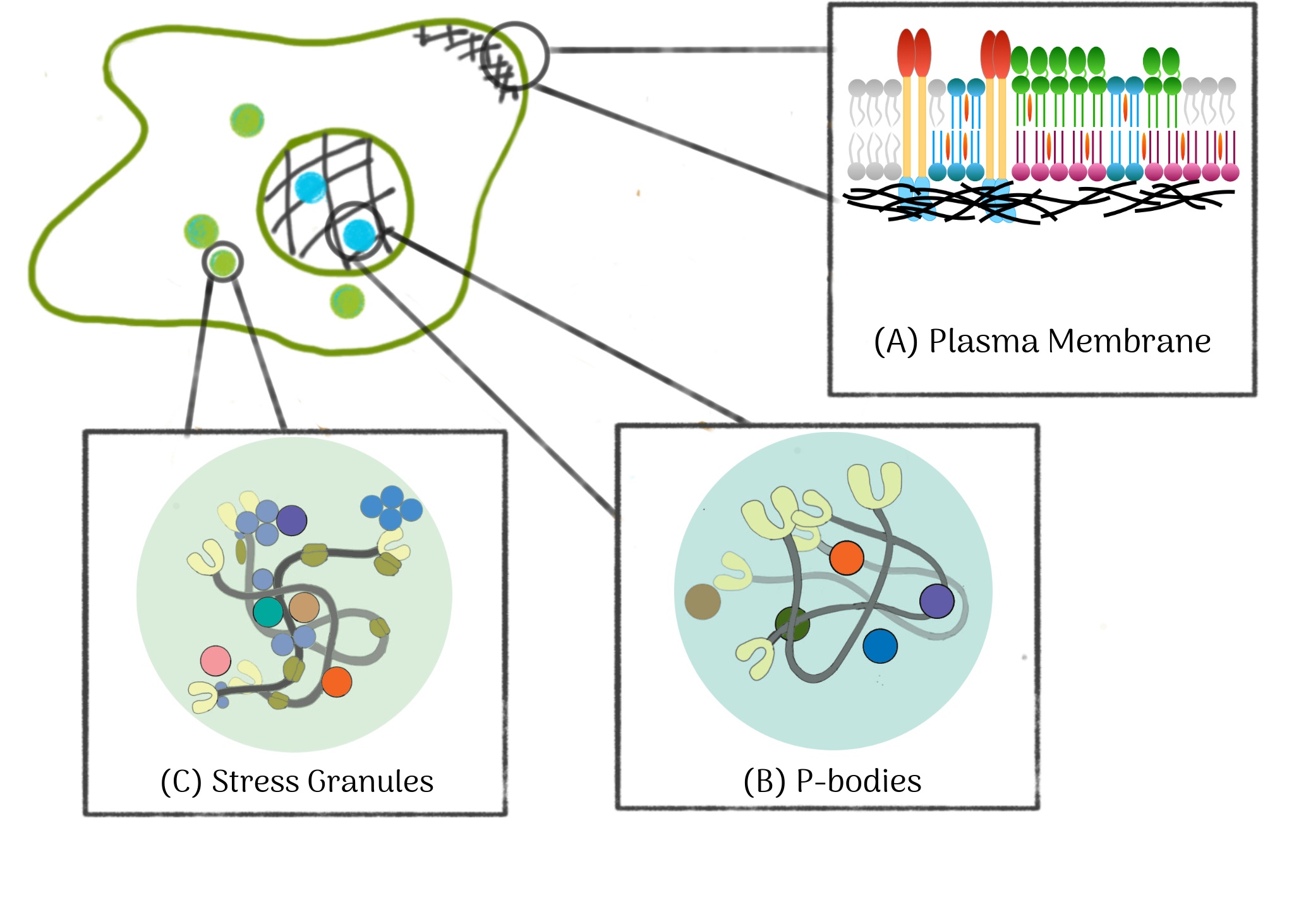}
\caption{Schematic of a typical metazoan cell highlighting candidate regions that exhibit nonequilibrium  phase segregation of composition at mesoscales - (A) The plasma membrane, juxtaposed with a layer of cortical actomyosin, is subject to active stresses which drive lipid and protein mesoscale segregation on the 2d membrane. (B-C) Nonequilibrium chemical processes have also been suggested as playing a role in liquid-liquid phase segregation in the 3d cell interior, such as in (B) P-bodies within the nucleus and (C) stress granules within the cytoplasm. }
\label{fig:schema}
\end{figure}

A striking example is the molecular organisation at the cell membrane at physiological temperatures~\cite{Edidin2003,Lingwood2010}. The components of the cell membrane are subject to fluctuating active contractile stresses from the thin cortical layer of actomyosin that adjoins it~\cite{Jacobson2019,Kripa2012}~(Fig.\,\ref{fig:schema}). We have proposed that these active stresses drive contractile flows that give rise to a lateral (2d) organisation of multiple species of lipids and proteins over different scales~\cite{Kripa2012,Raghupathy2015,MadanSatyajit2014}. 
This nonequilibrium organisation finds experimental  support from a variety of high resolution studies such as FRET~\cite{Kripa2012,Raghupathy2015}, super-resolution microscopy~\cite{Zanten2009} and 
EM~\cite{Plowman2005}. More recently we have investigated how such  nonequilibrium active stresses can give rise to a novel kind of mesoscopic phase segregation on the cell membrane
at physiological temperatures, which is larger than the putative equilibrium liquid-order phase transition temperature~\cite{Amit2016,Suvrajit2021}.
In addition to such motor-cytoskeleton complexes, localised chemical reactions at the cell surface described by reaction-diffusion-advection equations, can also generate mechanical or chemical stresses (via a nonequilibrium chemical potential) that may drive mesoscale segregation~\cite{Abhishek2011,Groves2019}.

Similarly, stress granules~\cite{Brangwynne2013,Parker2016} and 
P-bodies~\cite{Hyman2014}, belong to 
a growing class of membraneless organizations of specific biopolymers and proteins in the form of liquid-liquid condensates in the cell interior (Fig.\,\ref{fig:schema}).
While a variety of equilibrium mechanisms have been proposed for these assemblies (see~\cite{Berry2018} for a recent review), there have been recent suggestions that some of these condensates could be 3d realisations of active segregation~\cite{Julicher2019}.

In this paper, we combine some of our earlier studies on the active organisation and segregation of cell membrane composition~\cite{Kripa2012,Abhishek2011,Kripa2016,Kabir2017}, with newer insights from our~\cite{Amit2016,Suvrajit2021} and other studies~\cite{Cates2014,Cates2018},  to contribute to the general understanding of the dynamics of nonequilibrium phase segregation driven by activity. In doing so, we will highlight those aspects of the active segregation dynamics that are fundamentally different from equilibrium phase segregation.
For specificity, we will focus on the segregation of lipids on the plasma membrane of living cells, but our formalism has a more general scope. 

Giant unilamellar vesicles (GUV) with multi-component lipids, e.g., DOPC+SM+Cholesterol, and giant plasma membrane vesicles (GPMV) undergo an equilibrium liquid order\,-\,liquid disorder phase transition when quenched to  sufficiently low temperatures, $T_c=24.6^{\circ}$C for DOPC+SM+Cholesterol~\cite{Webb2010} and $T_c$ is typically below 
$25^{\circ}$C for GPMV~\cite{Baumgart2007}. Reference~\cite{Baumgart2007}  emphasizes that at 
$37^{\circ}$C, the GPMV membranes are almost entirely in the mixed, homogeneous phase.
Studies on the dynamics of coarsening of the lipid domains at $T<T_c$, subsequent to the initial linearly unstable growth regime, confirm the standard Cahn-Hilliard growth 
$R(t) \sim t^{1/3}$ driven by chemical potential gradients, with a prefactor that depends on the interfacial tension.
Our understanding of the physics of coarsening in these few-component ``artificial'' lipid systems is quite mature and summarized in the classic review of~\cite{Bray2002}.




On the other hand, 
the physics of segregation in the cell membrane differs fundamentally from that in artificial GUVs and GPMVs, because the cell membrane is additionally subject to a variety of nonequilibrium forces, primarily fluctuating active contractile stresses arising from a dynamic actomyosin cortex~\cite{Suvrajit2015,MadanSatyajit2014}. This has been the subject of discussion in several papers written over the years~\cite{Kripa2012,Abhishek2011,Kripa2016,Raghupathy2015,Kabir2017,Amit2016} and revisited in the current paper.
We will see that 
activity drives the system to a
nonequilibrium steady state with distinct features both in the dynamical approach to the steady state and the nature of fluctuations in the steady state. These features depend on how  activity enters into the dynamics of coarsening; specifically, on whether the segregating components act as {\it passive scalars} (that do not affect the agencies of activity, i.e., actomyosin) or {\it active scalars} (that act back on the agencies of activity).
This classification was introduced 
 in~\cite{Kripa2012,Abhishek2011,MadanSatyajit2014} and the analysis of active scalars initiated  
 in~\cite{Kabir2017}. More fundamentally, as has been highlighted 
in~\cite{Cates2014,Cates2018}, we will see that it is the 
breaking of time reversal symmetry (TRS) at the microscopic level  that is at the root of the profound differences between active and equilibrium phase segregation.

In earlier studies, we have investigated the effects of fluctuating active stresses
on the dynamics of clustering using the equations of active hydrodynamics~\cite{Kripa2012,Kripa2016,Kabir2017} and agent-based brownian dynamics simulations~\cite{Abhishek2011,Raj2021}.
In~\cite{Amit2016,Suvrajit2021}, we have studied the active segregation using a description based on a kinetic Master equation, which we solve by Monte Carlo simulations.
Finally, in~\cite{Kabir2017}, we have looked at the clustering of active advective scalars using an active hydrodynamics approach.
In this paper we formulate a new approach based on a time dependent Landau-Ginzburg dynamics starting from an active version of the Flory-Huggins theory~\cite{doi-edwards}, to study a hierarchy of active segregation models for a binary fluid (lipids) subject to active contractile stresses. We also report some new results on the relevance of active noise in the dynamics of coarsening
using the kinetic Master equation approach.

\section{Active Flory-Huggins theory}
\label{sect:AFH}

Consider a 2D surface (cell membrane) comprising two molecular components A ($lo$-lipids) and B ($ld$-lipids), whose areal densities are denoted by $\rho_A$ and $\rho_B$. A simple free energy functional describing its equilibrium configurations is given by,
\be
\label{eq:free1}
    F[{\rho_{A}},{\rho_{B}}]=\int \sum_{\alpha, \beta \in A,B}{\rho_{\alpha}}(r) v_{\alpha\beta}(r-r') {\rho_{\beta}}(r')\,dr\,dr'+T\int\sum_{\alpha \in A,B}\big[{\rho_{\alpha}}(r)\ln {\rho_{\alpha}}(r)-{\rho_{\alpha}}(r)\big] \,dr
\ee
where $v_{\alpha \beta}(r-r')$ is the non-local 2-body interaction. For  short-range interactions, one can expand it locally to obtain,
\be
\label{eq:free2}
    F[\rho_{A},\rho_{B}]= \frac{1}{2} \int\bigg( v_{\a\b}\, \rho_{\a} \rho_{\b}
   + c_{\a\b}\, \nabla \rho_{\a} \nabla\rho_{\b}\bigg)\,dr
 +T\int\sum_{\alpha \in A,B}\big[{\rho_{\alpha}}(r)\ln {\rho_{\alpha}}(r)-{\rho_{\alpha}}(r)\big] \,dr + \ldots
\ee

It is convenient to express this as a Flory-Huggins free energy functional~\cite{doi-edwards}, written in terms of the volume fraction of component A, $\phi=\rho_{A}/\rho$, together with 
$\rho = \rho_{A}+\rho_{B}$, which we take to be incompressible,
\be
\label{eq:flory}
     F[\phi]=T\int\bigg(-\chi(T,\rho)\, \phi^2+\kappa(T,\rho)\,(\nabla {\phi})^2+ 
     \mu(T,\rho) \,\phi+ \phi \ln{\phi}+(1-\phi)\ln({1-\phi}) + \ldots \bigg)\,dr
\ee
where $\chi(T, \rho)=\frac{(2v_{AB}-v_{AA}-v_{BB})}{2T}$, $\mu(T, \rho)=\frac{(v_{AB}-v_{BB})}{T}$ and $\kappa(T, \rho)=\frac{c_{AA}+c_{BB}-2c_{AB}}{2T}$.

We now describe the dynamics of segregation in terms of the hydrodynamic variables --
the volume fraction $\phi$ and the individual velocities ${\bf v}_{\a}$, where the barycentric hydrodynamic velocity,
  \begin{equation}
  \label{eq:hydro}
     {\bf v}=\phi \v_{A}+ (1-\phi) \v_{B}
 \end{equation}
 is incompressible, $\nabla \cdot {\bf v} = 0$.
The volume fraction $\phi$ of component A, obeys the usual continuity equation,
  \be
  \label{eq:contphi}
     \frac{\partial \phi}{\partial t} = - \nabla \cdot \left({\phi \v_{A}}\right) = \nabla \cdot \left({(1-\phi) \v_{B}}\right)
 \ee
The dynamics of the velocities ${\bf v}_\a$ is driven by exchange between the species and local stresses acting on the individual species~\cite{Tanaka2012}; assuming stationary flows, 
   \bea
 \label{eq:velphifastA}
      \gamma(\v_{A}-\v_{B})  & = & -\frac{1}{2}\nabla \cdot {\mathbf \Pi}_A  +\nabla \cdot {\mathbf \Sigma}_A + \phi \nabla p\\
     - \gamma(\v_{A}-\v_{B})   & = &  -\frac{1}{2}\nabla \cdot {\mathbf \Pi}_B + \nabla \cdot {\mathbf \Sigma}_B + \left( 1-\phi\right) \nabla p 
      \label{eq:velphifastB}
 \eea
 where the osmotic pressure is partitioned between the two components, 
 \be
\nabla \cdot  {\mathbf \Pi}_A = \phi \nabla \frac{\delta F}{\delta \phi}\, ,
\label{eq:osmotic}
 \ee
(similarly ${\mathbf \Pi}_B$) derivable from the Flory-Huggins free energy Eq.\,\ref{eq:flory} and the total stress tensor ${\mathbf \Sigma}$ is partitioned between the two components ${\mathbf \Sigma}_\a$~\cite{Tanaka2012}, a combination of dissipative and active stresses. The hydrodynamic pressure $p$ can be eliminated by using the incompressibility condition on the hydrodynamic velocity ${\bf v}$. 

The stress tensor $\mathbf \Sigma_\a$ for the two component fluid membrane subject to active contractile stresses arising from a coupling of cortical actomyosin to component A alone, is given by,
\bea
\label{eq:stress}
\mathbf \Sigma_\a & = & \mathbf \Sigma_\a^{(diss)} + \mathbf \Sigma^{(act)} \delta_{\a,A} \\
\mathbf \Sigma_\a^{(diss)} & = & \eta_\a^{(s)} \left(\nabla {\bf v}_\a + (\nabla {\bf v}_\a)^T
\right) + \eta_\a^{(b)} \left(\nabla \cdot {\bf v}_\a\right) \mathbb{1} \\
\mathbf \Sigma^{(act)} & = & \zeta_1(\rho_m,c)\, \mathbf{n n} + \zeta_2(\rho_m,c)\, \mathbb{1}
\label{eq:activestress}
\eea
where, the dissipative stress $\mathbf \Sigma_\a^{(diss)}$ is written in terms of the shear and bulk viscous stresses of the individual components, and the active stress $\mathbf \Sigma^{(act)}$, with $\zeta_1$ and $\zeta_2$ being negative for contractile stresses,
is expressed in terms of the density of myosin $\rho_m$ and the density $c$ and polarization $\bf n$ of actin filaments in the actomyosin cortex, for which we need to provide the dynamical equations~\cite{Kripa2012,Kabir2017}.
Since the dense actomyosin cortex is a momentum sink, one must in principle include a TRS violating contribution proportional to $\rho_m \,\n$, to
the right hand side of Eq.\,\ref{eq:velphifastA}.
These, together with Eqs.\,\ref{eq:hydro}-\ref{eq:activestress}, will form a closed set of equations that describe the segregation of a 2-component fluid membrane subject to active stresses.

One could imagine that the inter-particle interaction between the A-components could be modulated by the presence of a third component which is driven by the active contractile stresses. For instance, as shown in~\cite{Suvrajit2021}, the lower leaflet PS is such a component, but it could as well be the myosin density $\rho_m$. This would lead to a change in the local interaction parameter
$\chi(r) = \chi_0(T) - \chi_1(T) \rho_{m}(r)$, which 
would have the effect of locally enhancing the tendency for the A-component to segregate, if myosin ($\rho_m$) or PS were to accumulate.

 In what follows, we will 
 ignore the hydrodynamic velocity $\bf v$ in the overdamped dynamics of the 2d membrane embedded in 3d; this allows us 
to eliminate the pressure $p$ to get,
  \be
 \label{eq:velphifastelim}
      \v_{A}-\v_{B}  =    \frac{\left( 1-\phi\right)}{\gamma} \left[-\nabla \cdot {\mathbf \Pi}  - \frac{\phi}{1-\phi}\nabla \cdot {\mathbf \Sigma}_B + \nabla \cdot {\mathbf \Sigma}_A + \Gamma \, \rho_m\,\n \right]\, .
 \ee

It is convenient to go from the Active Flory-Huggins theory outlined above to an active Landau-Ginzburg formalism, written in terms of the segregation order parameter $\Phi \equiv 2 \phi - 1$, leading to the dynamical equation, 
    \be
    \label{eq:activeLG}
     \frac{\partial \Phi}{\partial t}=  \nabla \cdot \bigg( M(\Phi) \nabla \frac{\delta F}{\delta \Phi} - \frac{M(\Phi)}{\left(1+\Phi\right)} \bigg[ \nabla \cdot {\mathbf \Sigma}_A - \Gamma \,\rho_m \, \n \bigg]\bigg)
 \ee
 where we expect the relative exchange mobility $M$ between the two components to be proportional to $\left( 1-\Phi\right) \left( 1+\Phi\right)$. This reduces to the standard Model B dynamics~\cite{Halperin1977}  in the absence of active stresses.
 
When the dynamics of 
 $\Phi$ depends on the agencies of the active stress, but not vice versa, these set of equations describe the segregation 
of a passive scalar driven by fluctuating contractile active stresses. On the other hand, when the 
 agencies of the active stress in turn depend on $\Phi$, the resulting set of equations describe the segregation of an
 active scalar driven by fluctuating contractile active stresses.
 In Section III, we will take up  
 the active segregation of passive scalars, while in 
 Section IV, we will briefly discuss active segregation of active scalars, and show its relation to recent work on Active Models B and B$+$~\cite{Cates2014,Cates2018}.
 
\section{Active Landau-Ginzburg Models for segregation of passive scalars}
\label{sect:ALG}

The Active Flory-Huggins theory outlined above offers a systematic framework to arrive at a hierarchy of active Landau-Ginzburg models, each of which highlights a different aspect of the influence of activity on segregation dynamics.
Starting with the full dynamical theory in $\Phi$ and $\rho_m, c, \bf n$ (actomyosin variables), we systematically prune variables by declaring them to be    ``fast'' -- this implies that 
the theories higher up in the hierarchy contain all the features of those lower down. 

\subsection{Active Landau-Ginzburg Model 1}
\label{subsect:ALG1}
The dynamical equations for actin and myosin aggregates in the 2d cortex follow from~\cite{Kripa2012, Kabir2017} and many other sources, e.g., see~\cite{Marchetti2013} for the early references. These equations include contributions from active translational and rotational currents which in turn arise from active forces and torques.
The concentration of 
actin concentration obeys a continuity equation, 
${\pt c}=-\nabla \cdot {\bf J}_c$, with an
actin current given by 
 \be
 \label{eq:actinconc}
       {\bf J}_c = -{\bf D}_c  \nabla c +  
       c \, {\bs \gamma}\, \nabla \cdot \mathbf \Sigma^{(act)} +
       v_{0}(\rho_m)c{\bf n}
   \ee     
     where the first term is the diffusive flux with diffusion matrix ${\bf D}_c$, the second term is the active flux derived from active stress 
     with 
     active translational mobility ${\bs \gamma}$, and the last term is the flux from the active ``body force'' (a TRS violating term that is present because the cortex dissipates momentum via friction).

     The dynamics of the density of myosin aggregates has similar contributions to the flux, and in addition, includes a turnover from binding ($k_b$) - unbinding ($k_u$),
 \be
 \label{eq:boundmyo}
       \pt \rho_{m}=-\nabla \cdot \left( -{\bf D}_m \nabla \rho_m +  {\bf L}\, \nabla \cdot \mathbf \Sigma^{(act)} +v_{0}(\rho_m)\rho_m{\bf n}
      \right) +k_{b}(c)
       -k_{u}(\rho_{m})    
 \ee

 The dynamics of the polarization of the active actin filaments is given by
  \be
  \label{eq:orien}
       D_t \n  = K_{1}\nabla^2 {\bf n}+K_{2}\nabla(\nabla\cdot {\bf n})+
       {\bs \zeta}\, \nabla \cdot \mathbf \Sigma^{(act)}
 +\left(a(c-\bar{c}) - b \vert {\bf n}\vert^2\right)\, {\bf n}
 \ee
 where the differential operator $D_t$ includes the active self-advection~\cite{TonerTu,Marchetti2013}, 
 $K_{1}$ and $K_{2}$ are the Frank coefficients associated with splay and bend distortions, respectively, which in principle can depend on $\rho_m$, and the last term in the bracket ensures that the magnitude of $\vert {\bf n}\vert\approx 1$, with 
 $\bar{c}$, being the Onsager concentration of the polar filaments above which they become orientationally ordered~\cite{Gennes1993}.  
 
\begin{figure}
\centering 
\subfloat{\includegraphics[width=0.7\textwidth]{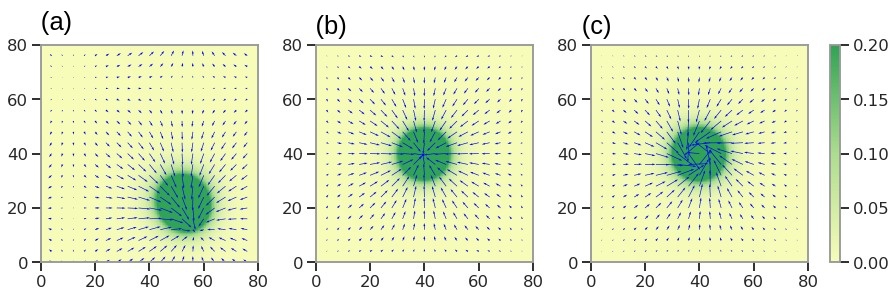}}
\newline
\subfloat{\includegraphics[width=0.7\textwidth]{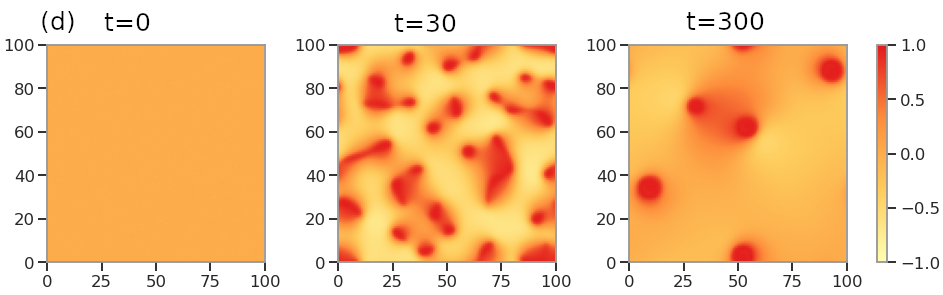}}
\newline
\subfloat{\includegraphics[width=0.7\textwidth]{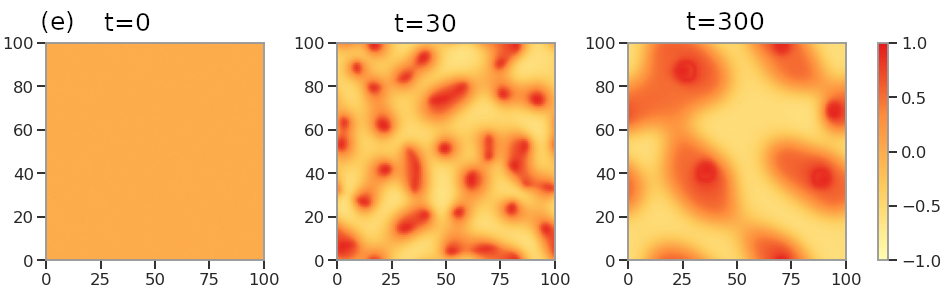}}
\newline
\subfloat{\includegraphics[width=0.7\textwidth]{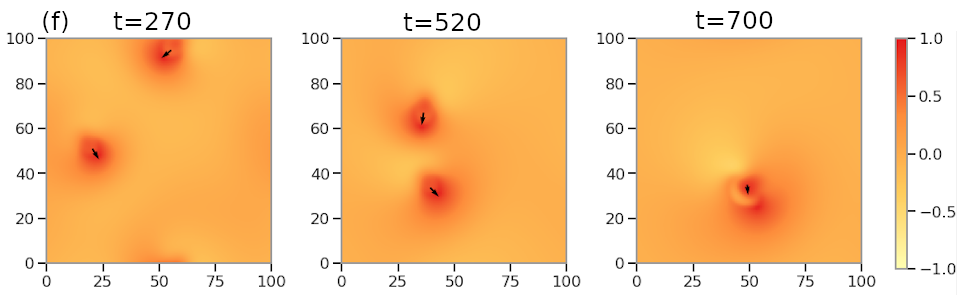}}
\newline
\caption{(a-c) Contractile clusters with definite filament orientation (arrows) and concentration profiles (heat map) emerge as an instability of the homogeneous, isotropic state by numerically solving Eqs.\,\ref{eq:actinconc}-
\ref{eq:orien} on a 2d grid for different values of the Frank coefficients $K$ and active parameter $\zeta$ :  (a) moving Polar cluster ($K = 3$, $\zeta = 30$), (b) stationary Apolar cluster ($K=1$, $\zeta = 30$) and (c) rotating Spiral cluster ($K=1$, $\zeta = 10$).  (d-e) Coarsening dynamics
driven by active contractile stresses at (d) $T>T_c$ and (e) $T<T_c$. (d) Spatial map of segregation parameter $\Phi$ (colour bar) shows coarsening of a symmetric mixture of A and B at $T/T_{c}=1.12$, 
starting from a homogeneous state at $t=0$ leading to the formation of mesoscale domains of the A-component at late times ($t=300$).
(e) Spatial map of $\Phi$ at $T/T_{c}=0.84$, shows coarsening into mesoscale domains of the A-component at late times ($t=300$). 
(f) Phase segregated domains at $T>T_{c}$ showing non-reciprocal features in the dynamics of propulsion and growth, in the regime which seeds polar clusters. Nonreciprocity is apparent during the propulsive movement of domains of component A (arrows), which shows a phoretic reorientation of the direction of propulsion of one of the domains prior to coalescence. The length and time scales are set by the filament translational and rotational diffusion coefficients~\cite{Kabir2017}. 
}
\label{fig:activeLG}
\end{figure}

The cortical fluid adjacent to the PM does not display global orientational order (dilute regime~\cite{doi-edwards}), even so, the active contractile stress fluctuations can drive large concentration fluctuations, such that locally $c(r,t) > \bar{c}$, leading to patterning instabilities which gets positively reinforced due to contractility. 
Following~\cite{Kabir2017}, we highlight 3 cases --
\begin{itemize}
\item When the active forces are larger than the active torques, the contractile stresses spontaneously create compact polar patterns  that are motile (Fig.\,\ref{fig:activeLG}(a)).
The size of the polar cluster is set by the interplay between filament diffusion and the contractile active force. The advection term $v_0$ distorts the shape of the cluster from circularity. It also sets the motility speed of the apolar cluster, which also depends on the asymmetric profile of actin and myosin in the cluster. The typical distance between these polar asters 
is given by the fastest growing mode
     $k^{-1}_{dist} \sim 1/{\sqrt \zeta}$.
     These compact contractile clusters of actin and myosin moving with a fixed speed correspond to travelling pulse solutions~\cite{Kabir2017}. 

\item When the active torques are larger than the active forces, the contractile splay instabilities give rise to compact apolar clusters that are stationary (Fig.\,\ref{fig:activeLG}(b)).
The size of the apolar cluster is again set by the interplay between filament diffusion and the contractile active force.

\item When the active force is comparable to the 
active torque, 
there is a spontaneous instability to the formation of spiral asters  that rotate with a local angular frequency proportional to $v_0 n_{\theta}(r)$
(Fig.\,\ref{fig:activeLG}(c)).

\end{itemize}
This patterning of concentration and orientation of actomyosin happens over a time scale
$1/\zeta$, with a fast build up of localised contractile stress that 
draws in other filaments as well as other neighbouring clusters leading to coalescence.

What effect does this spontaneous formation of contractile clusters have on the segregation dynamics of membrane components that interact with it? From Eq.\,\ref{eq:activeLG}, it is easy to see that this fast buildup of the active stress and the concomitant travelling pulse contributes to a driving force for the 
segregation and growth of the A component. This can be seen {\it even at $T>T_c$}, from a simple stability analysis about the homogeneous phase.

We perform a preliminary numerical analysis of Eqs.\,\ref{eq:actinconc}-\ref{eq:orien} and Eq.\,\ref{eq:activeLG} on a 2d grid, with parameters that lead to either polar or apolar clusters. Figures\,\ref{fig:activeLG}(d) and\,\ref{fig:activeLG}(e) represent snapshots of the segregation order parameter $\Phi$ profile for a symmetric mixture A-B, starting from the homogeneous phase, when
$T> T_c$ and $T < T_c$, respectively. Our numerical study confirms the instability towards segregation even when $T>T_c$. Unlike the  equilibrium segregation of symmetric mixtures, where one expects to see an evolving bicontinuous 
domain configuration, 
here we see 
the evolution of mescoscale domains, a consequence of the
 active terms in Eq.\,\ref{eq:activeLG} that break the $\Phi \to -\Phi$ symmetry.

Figure\,\ref{fig:activeLG}(f) shows snapshots of the order parameter $\Phi$ profile, following a quench into the
polar cluster phase. This shows the propulsive movement of domains of component A, and the phoretic reorientation of the direction of propulsion of one of the domains prior to coalescence. This nonreciprocal long-range sensing and rapid
coalescence of domains is a unique feature of the TRS violating active segregation dynamics~\cite{SSSR2014,Kabir2017,Golestanian,You2020,SRAS2020,Jyoti2021}. The TRS violation can be traced to the polarization 
$\n$, which is not a gradient of a scalar.
Such nonreciprocal effects of domain propulsion and growth would eventually lead to anomalous growth of domains.
The details of this coarsening dynamics and the influence of nonreciprocity on growth laws will be taken up later.


The segregation observed here bears some resemblance to the physics of motility-induced phase segregation (MIPS)~\cite{MIPS2015}.
The segregation of the passive scalars is driven by contractile flows into the core of the apolar clusters, where the density of the A-component gets larger. The mobility that multiplies the active part of the current in 
Eq.\,\ref{eq:activeLG}, is $\Phi$ dependent and gets smaller as $\Phi$ increases (i.e., as the concentration of the A-component increases).

\subsection{Active Landau-Ginzburg Model 2}
\label{subsect:ALG2}

So far we have taken both the actomyosin dynamics and the dynamics of segregation to be deterministic. We now look at the effects of noise, both thermal and active, which are inevitably present. Thermal noise that appears additively in  Eq.\,\ref{eq:activeLG} is known to be irrelevant in the coarsening dynamics, save for a finite renormalization of the interfacial tension~\cite{Bray2002}. What about active contributions to the noise, which arises as a consequence of stochastic turnover of actomyosin? 

To address this, we note that since the dynamics of the contractile agents is fast in the regime of high activity, it is useful to construct coarse-grained quantities, the scalar density and polarity of the compact contractile regions $\Omega$,
\begin{align}
\label{eq:coarse}
  \psi &= - \int _{\Omega} c \rho_m \nabla \cdot \n & {\bf p} &= \int _{\Omega} c \n 
\end{align}
In the limit of strong contractility ${\mathbf \zeta}$, 
the compact contractile clusters are apolar, and can be described by the coarse-grained field $\psi$ alone. 
The apolar cluster is described by a local orientation field that points radially inward
$\n \propto - \hat {\bf r}$ (Fig.\,2a)~\cite{Kripa2012,Kripa2016,Kabir2017}. 
This leads to an active {\it contractile} stress whose explicit form appears in the dynamics of $\Phi$ (Eq.\,\ref{eq:activeLG}).
We include the effects of athermal noise in the dynamics of the coarse-grained field $\psi$ -- ${\dot \psi} = k_{+} - k_{-}(\mathbf \Sigma^{(act)})\,\psi$ -- where the (stress dependent) turnover of actomyosin is
represented by a Poisson birth-death process~\cite{Basu2008,Abhishek2011,Kripa2012}.


In~\cite{Amit2016,Suvrajit2021}, we study this stochastic dynamics of active segregation using a kinetic Monte Carlo method. In~\cite{Amit2016} we  describe the active segregation in the simplest 2-component system in 2d driven by fluctuating active contractile stresses which captures the essential physics. 
The dynamics of the membrane components, subject to both equilibrium  and active forces, are described in terms of a Master 
equation for the time evolution of the probability distribution, $P(\{X^\alpha_i\}, \{{\bf x}_a\}, t)$, where ${\bf x}_a, a=1, \ldots n$, denotes the positions of the compact apolar clusters. We solve the Master equation 
using a kinetic Monte Carlo approach, where we specify the updates for the positions $\{X^\alpha_i\}$ of the membrane components and placement $\{{\bf x}_a\}$ of the active stress
events. While the  equilibrium exchange transitions  obey detailed balance, the advective moves on the A-component within the contractile regions and the 
birth-death moves of the contractile regions  do not (details in~\cite{Amit2016,Suvrajit2021}).


In~\cite{Suvrajit2021} we have generalised this theoretical framework to describe  the active segregation in a 5-component asymmetric bilayer containing {\it lo} components,  {\it ld} components and inner leaflet  PS. We show that it is the combination of actomyosin derived active contractile stresses and PS-mediated transbilayer coupling that drives the active segregation of the {\it lo} components. These predictions have been verified in high-resolution fluorescence-based experiments~\cite{Suvrajit2021}.


  \begin{figure}[h]
  \includegraphics[width=\textwidth]{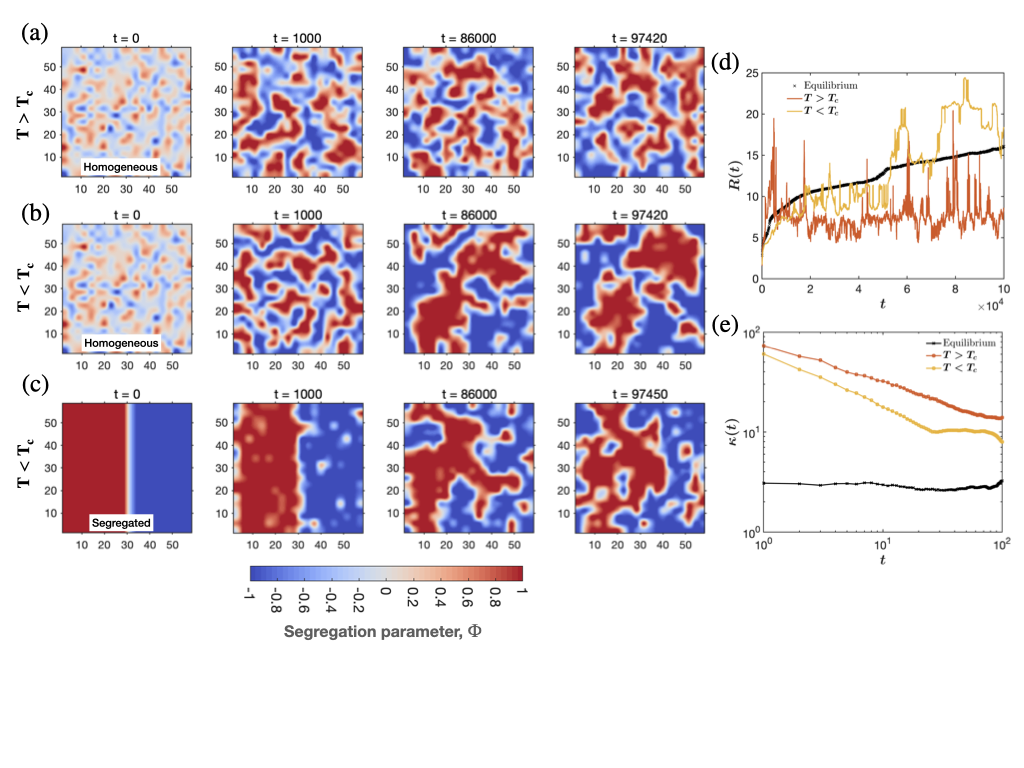}
  \caption{Coarsening dynamics
driven by fluctuating active contractile stresses at $T>T_c$ ($T/T_c = 1.06$) and $T<T_c$ ($T/T_c = 0.7$) obtained using a kinetic Monte Carlo simulation for a symmetric mixture of A and B described in the text.  Results shown here are for the same set of activity parameters - fraction of 2d space under the influence of active stresses is $0.3$, correlation length and time of active stresses is $\xi = 4$ and $\tau^{-1} = 0.13$ in units of particle size and particle diffusion rate, respectively.
(a) Spatial map of segregation parameter $\Phi$ (colour bar) 
shows coarsening at $T/T_c = 1.06$, 
starting from a homogeneous state at $t=0$ to the formation of mesoscale domains of the A-component (red) at late times ($t\sim 10^5$ MCS).
(b) Spatial map of $\Phi$ when $T/T_c = 0.7$, also shows coarsening into finite domains of the A-component at $t=1000$. 
(c) The fluctuating active stresses destabilise the fully segregated initial state at $T/T_c = 0.7$, leading to the formation of mesoscale domains of the A-component.  The nonequilibrium steady state obtained at late times is accompanied by large macroscopic fluctuations of a variety of statistical quantities. 
  (d) The time series of the domain size $R(t)$ obtained from the first zero of the correlation function $g(r)$ starting from the homogeneous state. The domain growth during equilibrium segregation at $T/T_c=0.7$ is consistent with $R(t) \sim t^{0.3}$~\cite{Amit2016}.  For the active segregation at $T>T_c$,  domain size appears to grow as $R \sim t^{0.25}$ before saturating to a finite value at steady state. The fluctuations of the domain size in the steady state are  large and abrupt, and correspond to large domains suddenly breaking up into smaller ones.
  (e) The large and abrupt macroscopic fluctuations of the domains  is a hallmark of {\it intermittency}, which shows up as a divergence of the time dependent kurtosis $\kappa(t)$ (4$^{th}$-moment of the domain size) as $t \to 0$ (in contrast, $\kappa(t)= 3$
  during equilibrium segregation). The large macroscopic fluctuations that characterise this nonequilibrium  steady state is an example of fluctuation dominated phase ordering (FDPO)~\cite{Amit2016}. }
  \label{fig:kineticMC}
\end{figure}


Figure\,\ref{fig:kineticMC} summarizes our results for the nature of the active phase segregation. We first note that as in Sect.\,\ref{subsect:ALG2}, even when $T>T_c$, an initial homogeneous state of a symmetric mixture of A and B, evolves to a state with mesoscale segregated domains of A at late times (Fig.\,\ref{fig:kineticMC}(a)). The instability of the homogeneous phase is driven by the apolar contractile clusters whose rapid turnover ensures that the segregation occurs over a large expanse of space. Similarly, Fig.\,\ref{fig:kineticMC}(b) shows
 snapshots of the coarsening dynamics when $T<T_c$; here too, instead of the usual bicontinuous domains seen during equilibrium segregation, one sees mesoscale domains of the A component. In Fig.\,\ref{fig:kineticMC}(c), we find that the fluctuating active stresses completely destabilise an initial prepared fully segregated configuration at $T<T_c$.

In contrast to equilibrium segregation dynamics where the dominant fluctuations are restricted to the interface, here the fluctuating active contractile stresses ensure that the fluctuations in the bulk too are significant. This is what drives the interface to be diffuse rather than sharp (as expected in equilibrium segregation) and the bulk to exhibit strong macroscopic fluctuations leading to 
breakup and re-formation of macroscopic structures. The interface dynamics is studied by looking at the behaviour of the dynamical structure factor $S(k) = \langle \Phi({\bf k},t) \Phi(-{\bf k},t) \rangle$ at large $k$ (small length scales). For sharp interfaces, $S(k) \sim k^{-3}$; this Porod behaviour characterises the growing domains during equilibrium segregation~\cite{Bray2002}. In contrast,
active coarsening  dynamics
shows  departures from  Porod behavior; the domains have diffuse interfaces and low interfacial
tension. 

On the other hand, bulk statistical quantities such as the integrated order parameter and domain size $R(t)$, show periods of quiescence followed by large changes over very short times (Fig.\,\ref{fig:kineticMC}(d)). This leads to intermittency in the steady state, a feature exhibited 
by many driven nonequilibrium systems such as turbulence; the fluctuating active stresses playing the role of stirring.
Intermittency, the sudden precipitous drops and rise of $R(t)$ in the active steady state, is captured by measuring the time dependent kurtosis of $R(t)$ (scaled fourth moment), which shows a divergence as $t \to 0$ (Fig.\,\ref{fig:kineticMC}(e)). As a result, the
steady state exhibits a continual breakup and reformation
of macroscopically large structures.

A striking implication of our results is that in contrast to equilibrium coarsening dynamics, where thermal fluctuations are irrelevant in the renormalization group sense~\cite{Bray2002}, in active  coarsening dynamics,
active stress fluctuations, parametrised by an active temperature related to the birth-death rates that drive the system to a distinct nonequilibrium steady state characterised by large macroscopic fluctuations (FDPO), are  relevant.
This fluctuation dominated
phase ordering (FDPO), is studied in great detail in~\cite{Amit2016} where we discuss its relation with other 
nonequilibrium models.
Activity destroys the very large domains
obtained in equilibrium phase segregation and makes them
more dynamic and intermittent.


\section{Active scalars = Active Model B/B$+$}
\label{sect:ActiveB}

Cell membrane molecules such as GPI-anchored proteins do not influence the dynamics of cortical actomyosin, and so it is appropriate to describe their actomyosin-dependent segregation using the active segregation models in Sect.\,III~\cite{Kripa2012,Raghupathy2015,MadanSatyajit2014,Zanten2009,Suvrajit2021}.  However, many signalling membrane proteins such as Integrin~\cite{Zanten2009} and Cadherin,
 or even myosin as studied in~\cite{Kabir2017}, can locally modulate the active stresses - these behave as {\it active scalars}.
 
 
 Thus taking component A to be an active scalar, we need to propose a dependence of the active contractile stresses on the segregation parameter $\Phi$.
  There are three distinct ways to do this:
\\

\noindent
(i)
We take 
 $\mathbf \Sigma^{(act)}$ to be purely isotropic and a function of $\Phi$ alone; it then contributes to the active renormalization of the osmotic pressure $\Pi^{(act)}(\Phi)$~\cite{Abhishek2011}. The effect would be indistinguishable from equilibrium segregation but for a shift in the critical temperature, leading to the formation of segregated domains when $T$ is greater than the equilibrium transition temperature $T_c$ (as in Sect.\,\ref{subsect:ALG2}).
 \\
 
 \noindent
(ii) We take 
 $\mathbf \Sigma^{(act)}$ to be isotropic, but now a function of $\Phi$ and $\nabla \Phi$, i.e. in addition to the above, the active osmotic pressure has, to lowest order, a contribution from $\Pi^{(act)}(\Phi, \nabla \Phi) \propto (\nabla \Phi)^2$. This would lead to a coarsening dynamics,
     \be
    \label{eq:activemodelB}
     \frac{\partial \Phi}{\partial t}=  \nabla \cdot \bigg( M \nabla \frac{\delta F}{\delta \Phi} + M \lambda  \nabla \left(\nabla \Phi\right)^2 \bigg)
 \ee
which could be interpreted as arising from
an active renormalisation of chemical potential,
 \be
 \mu = \frac{\delta F}{\delta \Phi}+ \lambda \left( \nabla \Phi \right)^2\, .
 \ee
This is equivalent  to the so-called Active Model B~\cite{Cates2014} and breaks TRS. 
 \\
 
 \noindent
(iii)
We now take the active contractile stress to be a general anistropic tensor of the form ${\mathbf \Sigma}^{(act)}  \propto - \zeta \triangle \mu\, {\boldsymbol n} \,{\boldsymbol n}$~\cite{hatwalne2004,Marchetti2013}. 
 On symmetry considerations, the influence of the segregation order parameter $\Phi$ on activity is given by a phoretic contribution $\n \propto {\boldsymbol \nabla} \Phi$ (as appears in~\cite{Kabir2017}) corresponding to an contractile stress
${\mathbf \Sigma}^{(act)}  \propto {\boldsymbol \nabla} \Phi \,{\boldsymbol \nabla} \Phi$. Conversely, a local patterning of the polarisation $\n$ generates irreversible fluxes of the order parameter $\Phi$. This leads to a coarsening dynamics of the form,
    \begin{equation}
    \label{eq:activeB+}
     \frac{\partial \Phi}{\partial t}=  \nabla \cdot \bigg( M(\Phi) \nabla \frac{\delta F}{\delta \Phi} 
     + {\mathbf L} \nabla \cdot \left({\boldsymbol \nabla} \Phi \,{\boldsymbol \nabla} \Phi\right) \bigg)
 \end{equation}
This is not a mere change in the chemical potential, but a renormalization of the current vector,
\be
{\bf J} = M \nabla \mu + {\mathbf L} \nabla \cdot \left({\boldsymbol \nabla} \Phi \,{\boldsymbol \nabla} \Phi\right)
\ee
equivalent to the TRS violating Active Model B$+$ dynamics~\cite{Cates2018}.
 
A detailed study of the dynamics of coarsening that shows striking nonreciprocal effects mediated through a long range wake in $\Phi$, similar to that described in~\cite{Jyoti2021}, will appear later.

 
 \section{Discussion}
 \label{sect:Disc}

 In this paper, we have described a framework to study nonequilibrium phase segregation in a binary mixture driven by active contractile stresses using an active version of the Flory-Huggins formalism. Though our description is more general, here we focus on compositional segregation in the plasma membrane, where
  several studies~\cite{Kripa2012,Raghupathy2015,MadanSatyajit2014,Zanten2009,Plowman2005} have shown that one of the primary driving forces behind lateral segregation of composition, both lipids and proteins, at the cell surface at physiological temperatures, 
 is the nonequilibrium active contractile stresses from the actomyosin cortex adjoining it. In applying our general formalism to this context,  
 we have brought together some of our past theoretical work on active clustering and segregation of passive and active scalars~\cite{Kripa2012,Abhishek2011,Kripa2016,Amit2016,Kabir2017}.
 
 In this paper, we have used both a hydrodynamic approach~\cite{Kripa2012,Kripa2016,Kabir2017} and a kinetic Monte Carlo simulation~\cite{Amit2016}. 
 The most striking results are (i) instability to segregation even at $T>T_c$, (ii) segregation into mesoscale domains for symmetric binary mixtures even when $T<T_c$,
 (iii) nonequilibrium segregated state characterised by macroscopic and abrupt fluctuations, (iv) propulsion and nonreciprocal features enroute to domain coalescence in systems where frictional dissipation dominates, and (v) fuzzy domain interfaces suggesting low interfacial tension.
 These studies have culminated in a recent proposal of an {\it Active emulsion}~\cite{Suvrajit2021} as a description of the mesoscale organisation of 
 lipids on the cell membrane at physiological temperatures, much larger than the putative equilibrium liquid-order phase transition temperature, and contingent on both actomyosin activity and a (lower leaflet) PS-mediated transbilayer coupling. The generality of our theoretical approach suggests that such {\it Active emulsions} could be a general description of mesoscale segregation in actively driven binary fluids and of mesoscale organisation of molecules driven by a combination of nonequilibrium stresses and thermodynamic forces, in a variety of cellular contexts.
Its immediate implications for local composition control in the cell make this viewpoint compelling.

\section*{Acknowledgements}
We would like to acknowledge the tremendous contribution of many past students and postdocs, in particular,
K. Gowrishankar, K. Husain, A. Chaudhuri, B. Bhattacharya and A. Polley, whose work over the years has culminated in our current proposal of active emulsions described here and in~\cite{Suvrajit2021}. We would also to thank our experimental colleagues, S. Mayor,
and S. Saha, with whom we have had years of active collaborations. We thank S. Ramaswamy for discussions on the manuscript.
 We acknowledge support from the Department of Atomic Energy (India), under project no.\,RTI4006, and the Simons Foundation (Grant No.\,287975), and computational facilities at NCBS. AD acknowledges support from the Centre for Theoretical Biological Physics at Northeastern University and the Discovery Cluster at Northeastern University.
MR acknowledges the award of JC Bose Fellowship from SERB-DST, India.

\end{document}